\documentstyle[preprint,aps]{revtex}
\input psfig
\newcommand{\psplot}[1]{ \hbox to\textwidth{\hfill
             \psfig{figure=#1.ps,width=4.0in,height=4.0in}\hfill} }
\begin{document}

\draft 
\tighten

\preprint{\vbox{\hbox{CLNS 96/1424\ \ \ \ \  \hfill}
                \hbox{UCSD/pth 96-19\ \ \ \ \  \hfill}
                \hbox{VPI-IPPAP-96-4\ \ \ \ \  \hfill}
                \hbox{July 25, 1996    \hfill}}}

\title{A Compact Gas \v{C}erenkov Detector with Novel Optics}
\author{
M.~Sivertz}
\address{
University of California, San Diego, La Jolla, California 92093}
\author{
B.E.~Berger,  R.D.~Ehrlich}
\address{
Cornell University, Ithaca, New York 14853}
\author{
J.~Bartelt, S.E.~Csorna, V.~Jain, and S.~Marka}
\address{
Vanderbilt University, Nashville, Tennessee 37235}
\author{
K.~Kinoshita, P.~Pomianowski}
\address{
Virginia Polytechnic Institute and State University,
Blacksburg, Virginia 24061}
\maketitle

\begin{abstract}

We discuss the design and performance of a threshold \v{C}erenkov
counter for identification of charged hadrons.
The radiator is pressurized gas, which is contained in thin-walled
cylindrical modules.
A mirror system of novel design 
transports \v{C}erenkov photons to photomultiplier tubes.
This system is compact, contains relatively little material, and has a 
large fraction of active volume.
A prototype of a module designed for the proposed CLEO~III
detector has been studied using cosmic rays.
Results from these studies show good agreement with 
a detailed Monte Carlo simulation of the module and indicate that
it should achieve separation of pions and kaons at the 2.5$\sigma$-3.0$\sigma$
level in the momentum range 0.8-2.8~GeV/c.
We predict performance for specific physics analyses 
using a GEANT-based simulation package.  

\end{abstract}

\section{Introduction}

The identification of stable charged hadrons 
at momenta above 1~GeV/c is difficult to accomplish,
particularly in a colliding beam environment where
spatial limitations constrain detector design.
This problem has received increased attention in the past few
years in the context of flavor-dependent
processes (e.g. $CP$ violation) in decays of $B$ mesons, where pions and
kaons must be identified with high efficiency and accuracy.
To study such processes in $e^+e^-$  collisions at the
$\Upsilon$(4S) resonance will require devices which can provide 
identification for particles
with momenta up to $\sim 4$~GeV/c.

Presented here is a design for a particle identification (PID) device
based on the \v{C}erenkov effect, working mainly in the
threshold mode with a radiator of pressurized gas.
The detector concept is generally applicable in situations
with moderately high rates and low particle multiplicities. 
It has been applied here to the proposed
CLEO~III detector\cite{cleoiii}, an upgrade to the CLEO~II
detector at the Cornell Electron Storage Ring, and
some of the detector parameters reflect our accommodation 
of constraints imposed by the existing system.
Specifically, the radial gap available for a PID device was fixed at
20~cm, between the main tracking chamber (at a radius of 82~cm) and the
calorimeter.
In addition, 
it was prescribed that the PID device should comprise no more than
0.12 radiation lengths of material, so as to have a minimal impact
on the performance of the CsI calorimeter.
We first discuss the overall approach used to arrive at a design,
including development of components and tests needed to establish
validity.
A description of a working prototype and tests with cosmic rays
follows.
Finally, results from these tests are used to tune a 
Monte Carlo simulation, which is then 
used to evaluate the performance of a full-scale system in 
the proposed CLEO~III detector\cite{cleoiii} for physics analyses 
in symmetric $e^+e^-$ 
collisions in the $\Upsilon$ region ($\sqrt{s}\sim$10~GeV).

\section{Detector Design}
\label{overview}
A threshold \v{C}erenkov counter has several advantages over other
PID devices.
Its readout can be simple and the number of channels few.
The response time is intrinsically fast, and timing and pulse height
calibration requirements are minimal.
These features make the offline software requirements rather simple.
Furthermore, such a device may be included in the trigger.

Design considerations are dominated by the need to maximize 
photon detection efficiencies,
because the absolute number of photons produced by \v{C}erenkov radiation
is relatively small.
It is also desirable to restrict the material
thickness of the detector and to minimize inactive volume within the
central region.
Other considerations include ability to operate in
high magnetic fields, robustness, cost, and personnel requirements.

As is described below, the index of refraction required for the
desired threshold limits the choice of radiator to pressurized gases
or silica Aerogel \cite{aerogel}.
We pursued the possibility of pressurized gas because it has better
transparency, particularly at shorter wavelengths, where the 
\v{C}erenkov spectrum is concentrated.
Another advantage of gas is that there is some flexibility to
vary the threshold through adjustments of the gas pressure.
The high transparency enables the design of a long cylindrical detector module,
in which radiated light is transported up to 2~meters before being detected.
The modules may thus be arranged to span the central
solid angle of a collider detector, and
are sufficiently large that the number of readout channels 
required is small, of order 100.
This system can be made conservatively, relying only on 
readily available components and proven technologies.
More specific aspects of the design are discussed below.

\subsection{Radiator}
\label{radiator}

In many detector systems, and specifically in this case, hadrons 
with momenta below 0.8~GeV/c are readily identified through 
measurements of specific ionization (dE/dx) in tracking devices,
so that the desired range of operation for an additional PID device 
is in the region above 0.8~GeV/c.
To provide good identification at 0.8~GeV/c 
we select an index $n=1.02$, which places the pion threshold at 0.70~GeV/c.
The kaon threshold for this index is 2.45~GeV/c, which is near the kinematic
limit for products of $B$ decay in the rest frame of $\Upsilon$(4S) events.

For \v{C}erenkov photons to be detected,
the radiator must transmit them, and
the photodetector must be sensitive to them.
For a system sensitive to photons with energies between
$E_{min}$ and $E_{max}$ (in eV), a particle traveling with speed $v=\beta c$ 
a distance $L$ (in cm) through a medium with refractive index $n$
produces
\begin{equation}
N_{\gamma} \approx 370 \sin^{2}\theta_C L (E_{max} - E_{min})
\label{eq:introone}
\end{equation}
detectable photons, where $\theta_C \equiv\cos^{-1}(1/\beta n)$ is 
the \v{C}erenkov angle.
The most effective device for detection of individual photons is still
the conventional photomultiplier tube, which maintains a 
high quantum efficiency for wavelengths below 600~nm and may detect
photons with wavelengths below 200~nm if the window is made of a medium 
transparent to them, such as quartz.
Sensitivity to shorter wavelengths is desirable, as
the \v{C}erenkov  spectrum is more concentrated in that region.
However, because many common impurities absorb strongly at
wavelengths below 200~nm, 
the practical range for detection is 200-600~nm (2.1-6.2~eV).
In this range, in a material with index $n=1.02$,
1~GeV/c pions radiate 30 \v{C}erenkov photons/cm, while 2.8~GeV/c 
pions radiate 60 photons/cm.  
If the detector system is able to count the number of photons, some
separation of pions and kaons can be maintained even above the kaon 
threshold; 
only 11 photons/cm are emitted by 2.8~GeV/c kaons.
Thus, pions and kaons at momenta in the
range 0.8-2.8~GeV/c produce \v{C}erenkov signals which should be 
readily distinguishable from each other.

The selection of \v{C}erenkov gases at our chosen index is limited,
as many either liquefy under pressure at room temperature
before reaching the target index or require undesirably high pressures.
Sulfur Hexafluoride (SF$_6$) satisfies our criteria.
At an absolute pressure of 20 atmospheres, its room-temperature
index of refraction is $n=1.02$.
Its inertness, non-flammability, and previous use as
radiator~\cite{Atkinson,Garwin} make it an attractive candidate for
this application.  
Ethane (C$_2$H$_6$) is an alternative which reaches the same index at
somewhat higher pressure but has a longer radiation length.
The higher pressure required, in combination with its flammability,
dissuaded us from pursuing ethane for this design.

For pathlengths up to 1~atm$\cdot$m (1~m pathlength at 1~atm),
SF$_6$ is known to have excellent transparency
to light with wavelengths above 190~nm\cite{Garwin}.
However, because this application requires good transparency over a
pathlength of 20~atm$\cdot$m,
we performed an independent measurement
using a McPherson 218 0.3~m scanning monochromator illuminated by
a deuterium lamp.  
The monochromator was made to scan through wavelengths from 200~nm to
350~nm.
The resulting beam traversed 1~meter of gas
contained in a pressure vessel with sapphire end windows and was detected 
in an EMI-Gencom RFI/B-214 solar-blind photomultiplier tube.
The anode current of the photomultiplier, which is proportional to the
intensity of the transmitted light, was converted to a DC voltage
by a Keithley~440 picoammeter, digitized by an AR-B3001 ADC plug-in
card (Acrosser Technologies, Inc.), and read out at 0.25~nm intervals.
The wavelength scanning and data acquisition of this system were
controlled by a personal computer (IBM AT286 equivalent).

Two grades of SF$_6$ were studied, instrument (99.99\% pure) and
ultra-pure (99.996\%).  
As a baseline reference we used
ultra-pure nitrogen (99.9999\%), which has a transparency 
comparable to vacuum.  
For a pathlength of 1~atm$\cdot$m, both grades of 
SF$_6$ were found to be as transparent as nitrogen above 200~nm, 
in agreement with \cite{Garwin}.  
Each gas scan was therefore normalized to its own 1~atm baseline rather 
than to nitrogen,
reducing substantially the systematic error from
variables such as changes in alignment between gas fills or 
drifting of electronic calibrations over time.

\begin{figure}
\centerline{\psfig{figure=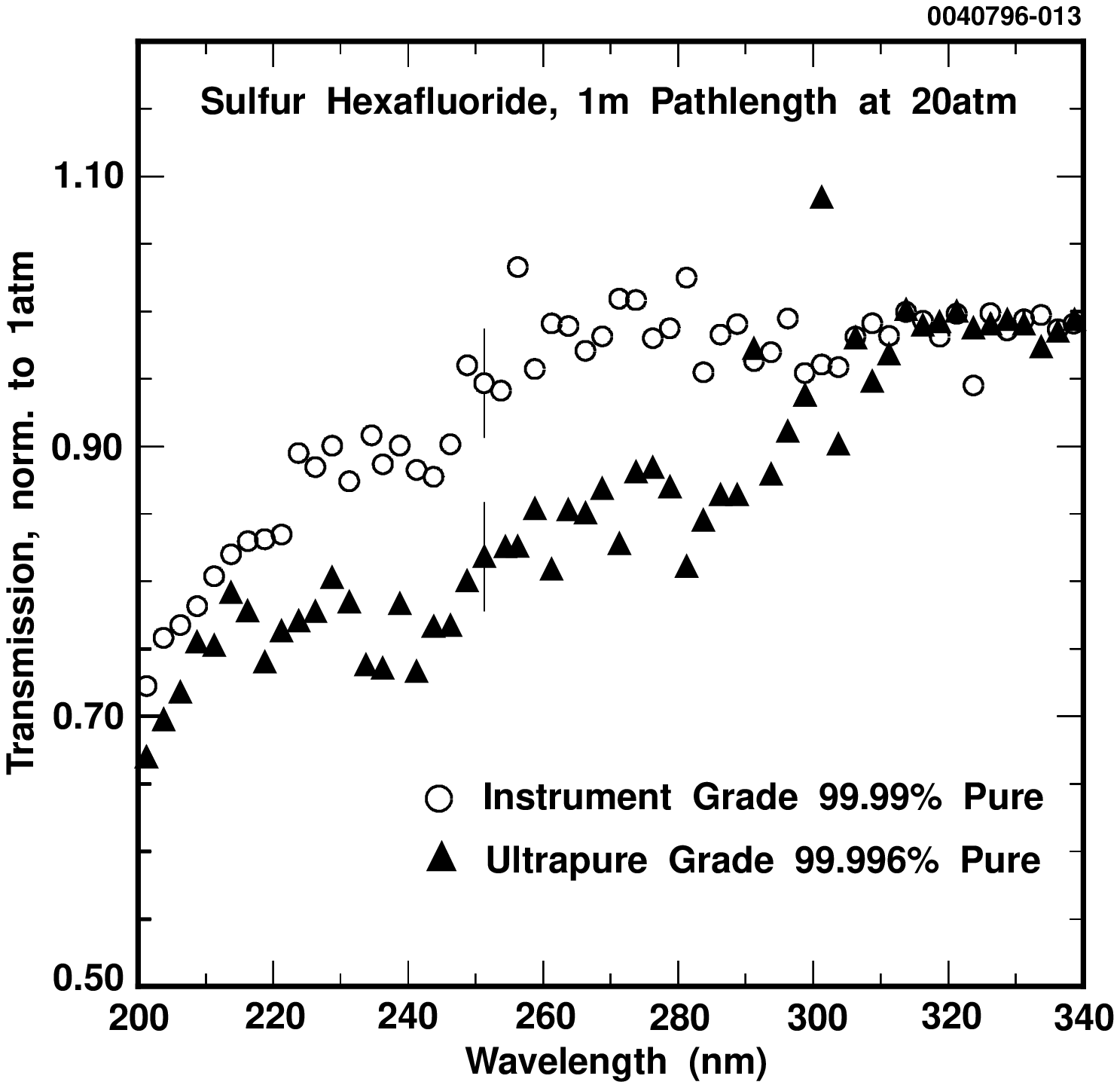,width=5.in}}
\caption{Transmission curves for two grades of $SF_6$ at a pressure of 
20~atm. Because of low light yields, each point has an error of $\sim$4\%,
which is shown for only two data points.
}
\label{fig:sf6_280}
\end{figure}

The gas vessel was initially filled with SF$_6$ at 
the maximum pressure delivered by the cylinder ($\sim$20~atm) and allowed to
stabilize~\cite{foot}. 
Scans were then taken at different pressures by
releasing gas successively until the ambient pressure was reached, at which
point a scan was taken to act as the baseline for that set of
measurements.
The resulting transmission efficiencies as a function of
wavelength are shown in Figure~\ref{fig:sf6_280}.
20~atm$\cdot$m of both instrument and ultrapure SF$_6$ were found 
to have transparencies comparable to 1~atm$\cdot$m nitrogen, down to 320~nm.  
However, at
shorter wavelengths, the instrument grade is more transparent than
the ultra-pure, confirming the results in \cite{Tomkiewicz}.  
More than
90\% of light down to 240~nm is transmitted by the instrument grade,
dropping to 72\% by 200~nm as Rayleigh scattering becomes more
prevalent.  
Although it is not known to this group why the lower
purity gives better performance, it has the fortuitous result of greatly
reducing the cost of the gas needed for this detector.

\subsection{Pressure module}

The pressure vessel is designed with a cylindrical shape to insure
the required mechanical strength with minimum wall thickness.
Within each module, photons are transported via reflections off 
mirrored surfaces to the tube ends, where they enter a
transparent window and are collected by photomultipliers.
The vessel itself should be composed of a strong, low-$Z$ material
to minimize the number of radiation lengths of material penetrated
by particles.

\begin{figure}
\centerline{\psfig{figure=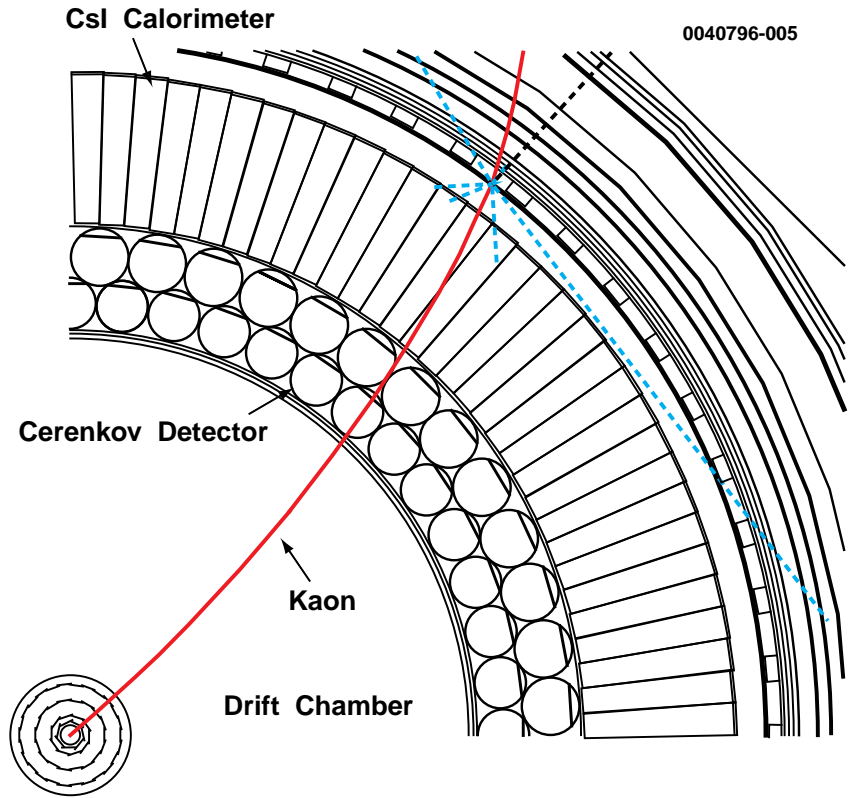,width=4.5in}}
\caption{Configuration of detector modules in a colliding beam experiment.}
\label{fig:system}
\end{figure}

In designing a device for CLEO~III we envision modules
with $\sim$10~cm diameter and 2~m length.  
Two close packed layers, arranged in a barrel, 
surround the interaction point just outside the central 
tracking chamber (Figure~\ref{fig:system}).
An economical option for the tube material is aluminum.
A 0.1~cm wall thickness is within the limits
prescribed for CLEO~III and is adequate from the
point of view of safety. 
The hoop stress for this design is about
1$\times 10^8$~Pa, while Al tubes of grade 7075-T6 will maintain their
strength to about 5$\times 10^8$~Pa, providing a five-fold safety factor.
Other materials considered include carbon fiber and beryllium.
These could reduce the material thickness but were
not studied fully.

\subsection{Optics}

The relatively small diameter and long transport distance
of the envisioned module place severe constraints on the optical
design.
To obtain good collection efficiency, photons should experience no more
than 4-5 reflections before reaching the photomultiplier, as
reflectivities in the ultraviolet region can be expected to be 
$\sim 80\%$ at best.
The simple option of placing a smooth cylindrical reflecting liner
in the pressure vessel works remarkably well for tracks which are
moderately angled toward the tube ends, at polar angles of 
65$^\circ$ or less ($|cos\theta |>0.42$).
We used a mylar sheet, 0.076~mm thick and coated with a thin film
of Al and a MgF$_2$ overcoating\cite{hyper}, which was rolled and placed inside the tube.
Measurements on one sample indicate that its reflectivity at 
normal incidence ranges  from 76--80\% at 200~nm to approximately 
90\% at 600~nm. 

\begin{figure}
\centerline{\psfig{figure=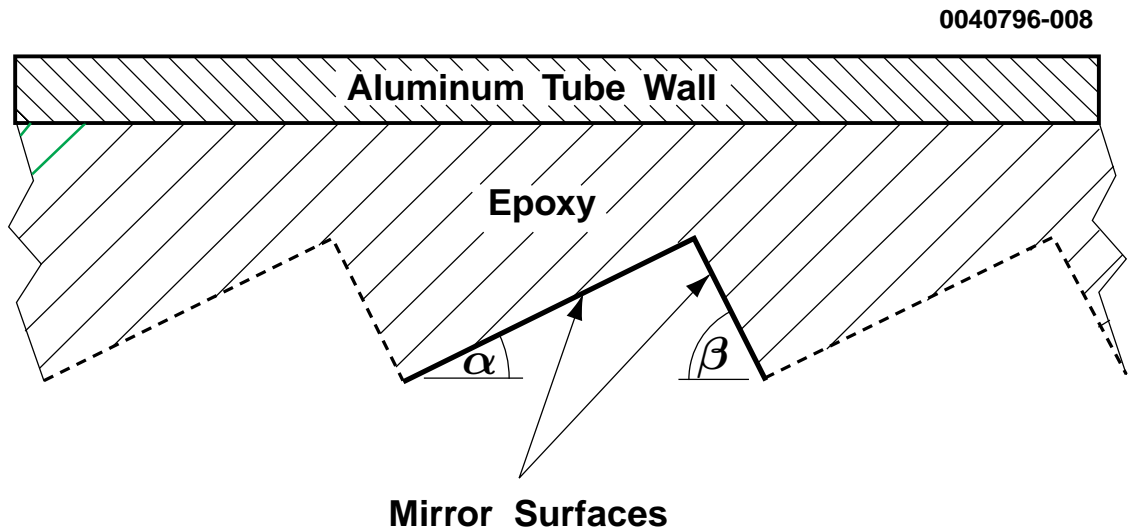,width=6.0in}}
\caption{Cross section of faceted mirror.  The nearer photomultiplier
is toward the right in the figure.}
\label{fig:grating}
\end{figure}

\begin{figure}
\centerline{\psfig{figure=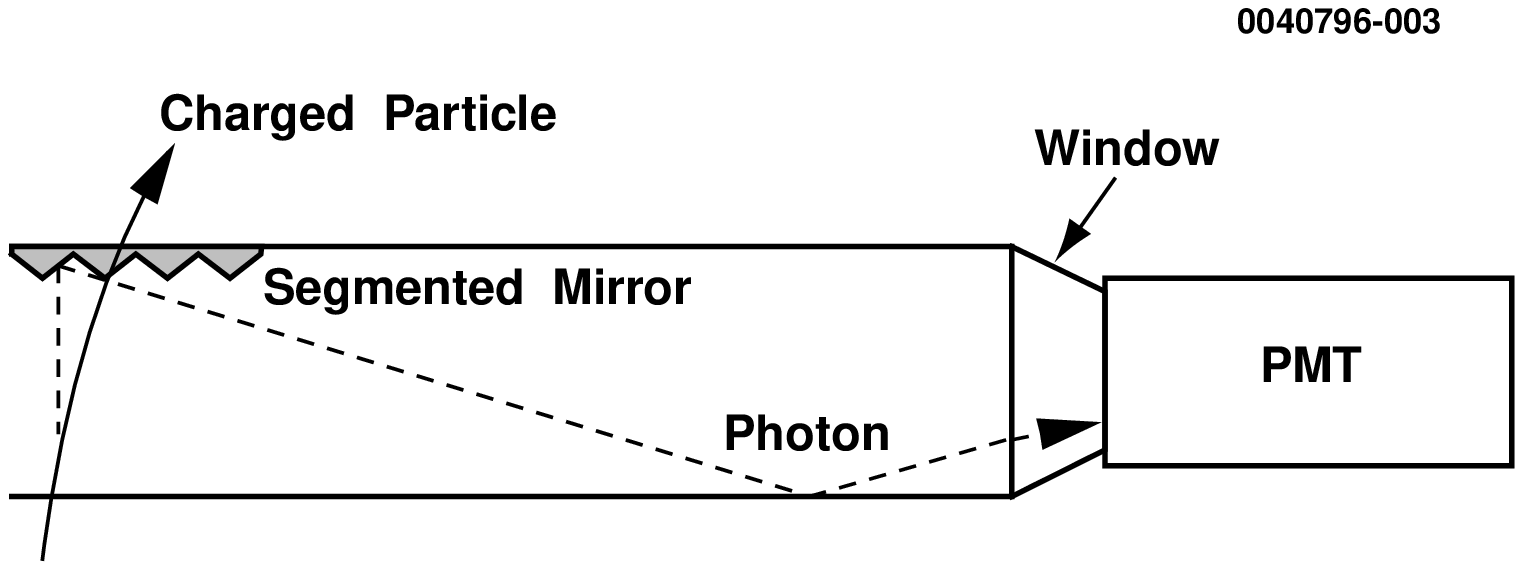,width=5.0in}}
\caption{Process for transport of photon from center of
detector module to photomultiplier.}
\label{fig:bounce}
\end{figure}

For tracks in the central region, at polar angles of 65$^\circ$
to 115$^\circ$, the simple cylindrical mirror does not transport
sufficient light.
Monte Carlo simulations show that the efficiency for transporting photons to
either end of the tube drops from 30\% at 65$^\circ$
to zero at 90$^\circ$.
To improve this situation we have developed a 
specialized reflector for the central region.
It is designed to redirect
a large fraction of the photons such
that they exit the region before encountering another mirror surface
and are then able to reach the end of the module with just one or
two additional reflections.
This reflector has a faceted surface with a sawtooth-shaped cross section  
(Figure~\ref{fig:grating}).
Each facet is a planar strip with a width of order 1~mm \cite{width},
oriented such that an incoming photon produced by a particle from
the interaction point is reflected toward the end
of the tube, as shown in Figure~\ref{fig:bounce}.
This faceted mirror system extends parallel to the beam line, 0.61~m
to either side of the interaction point
($z=-0.61$~m to $z=0.61$~m), is about 5 cm wide, and faces the beam line.

The optimal combination of facet orientations varies with the polar angle
of the incident particle.
Our design incorporates
ten different mirror sections, each with area 
$\sim 5\times 5$~cm$^2$ and a different 
combination of angles (Table~\ref{tab:angles}).
The facets reflecting photons toward the nearer phototube make an angle
$\alpha$ with respect to the $z$ axis, while those reflecting
photons toward the more distant PMT make an angle $\beta$, as shown
in Figure~\ref{fig:grating}.
The optimal values of $\alpha$ and $\beta$ as a function of $z$ depend 
on the tube length, its radius, the perpendicular distance to the 
interaction point, and the reflectivity as a function of wavelength.
A Monte Carlo simulation was used to find a set of discrete 
angles for the ten sections spanning $0.0$~m~$\le |z|\le$~0.61~m.

Once a photon reaches the end of the tube, it may need to be funneled to a smaller
area in order to be delivered to a photocathode;
affordable PMT's have typical photocathode diameters of 3.8 or 6.4~cm.
In addition, a PMT cannot survive a high pressure environment,
and must therefore be located outside the pressure vessel.
The light is delivered to the PMT
through a quartz window which serves as
pressure barrier, UV-transparent window, and light funnel.
For this test the window was made of Corning excimer grade quartz,
in the shape of a simple 
truncated cone, with an inner face slightly smaller than the diameter 
of the tube (9.7~cm),
an outer face matching the photocathode (6.4~cm diameter),
and a thickness of 3.8~cm.
The conical side wall was coated with aluminum.
This window
works in two ways: because of the change in index of refraction, rays
that come in at large angles of incidence refract to smaller angles,
thereby aligning their trajectories more parallel to the axis of the
tube; secondly, many of
the photons striking the mirrored conical wall are directed into the PMT. 
Monte Carlo studies of photon transport predict that such a
window is about 65\% efficient in transmitting photons from
the inner face of the window to the photocathode.

To minimize transmission loss,
the windows of the pressure module and the PMT must be coupled to each 
other by an index matching fluid. 
Using the monochromator apparatus described in section~\ref{radiator},
we compared transmission of light through the window with and without
each of four recommended coupling greases, General Electric Viscasil,
Oken 626A, Dow Corning Q2-3067, and Shinetsu.  
Of the four, only the General Electric Viscasil performed well.
It improved transmission through the window down to 220~nm
and gave diminishing efficiency at lower wavelengths, dropping to
80\% at 200~nm.
The other three did not transmit light
efficiently below 270nm (less than 40\% transmission).  

\subsection{Photomultipier}

In many experimental configurations, the PID detector is located
in a strong magnetic field.
In such cases, conventional photomultiplier tubes cannot function, and
it is necessary to use a specialized photodetector 
which is able to operate in such an environment.  
We tested one such PMT, a Hamamatsu R3386 ``fine mesh'' tube
\cite{fmpmt}, to evaluate
its suitability for this application. 

For our \v{C}erenkov detector the primary concern is resolution, particularly
at low light levels;
loss of gain may in principle be recovered with 
additional amplifiers.
Signal resolution in conventional PMT's is determined mainly by photoelectron 
counting statistics, as
well as the quantum efficiency of the photocathode.
In fine mesh tubes the resolution is complicated because
the dynodes are partially transparent ($\sim$50\%), yielding 
anomalously shaped distributions  at low intensities.
In fact, a single photoelectron signal will not show a well-defined peak,
due to this partial transparency\cite{enomoto}.

To evaluate the Hamamatsu R3386, we set up a test system to deliver
calibrated light pulses to the PMT and then to measure the gain
and resolution of the output.
A BNC light pulser was used to deliver up to several hundred photons, and
pulse height spectra were recorded using a LeCroy qVt
and a Canberra multichannel analyzer.
The light output of the pulser was calibrated using a
Hamamatsu R2059 ``photon counting'' photomultiplier tube.
The fractional width of the R2059 was found to be ($1.01\pm 0.02)N^{-{1\over 2}}$,
where $N$ is the number of photoelectrons, in agreement with
expectations for a ``photon counting'' PMT.

\begin{figure}
\centerline{\psfig{figure=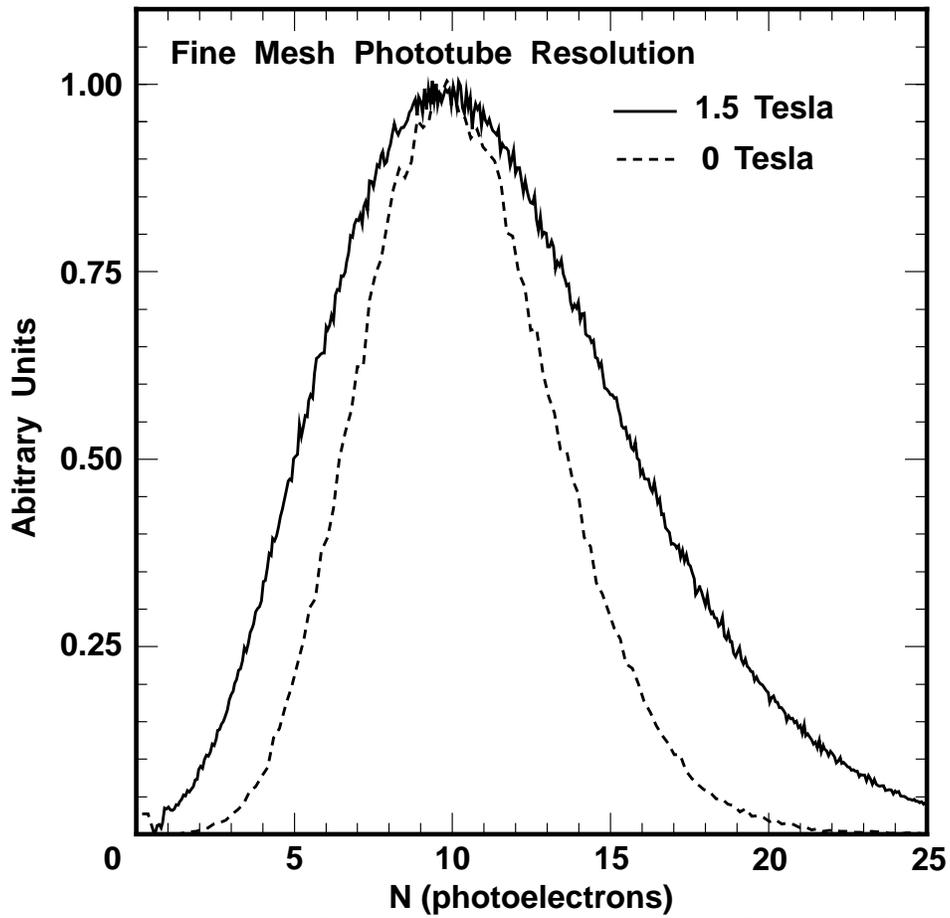,width=5.0in}}
\caption{Signal distributions for R3386 fine mesh
phototube in 0~T and 1.5~T magnetic fields, for light pulses
producing an average of 10 photoelectrons.
Both signals are normalized to a mean of 10 photoelectrons.}
\label{fig:pmt}
\end{figure}

The gain and resolution of the R3386 were then measured
in and out of the CLEO~II solenoid's 1.5~Tesla magnetic field.  
For this test the axis of the PMT was oriented parallel to the field.
The high voltage on the R3386 was varied in the range 1200-2000~V at 0~T,
and 2500-2700~V at 1.5~T, such that all measurements
could be carried out within the same range of signal heights.
The gains and widths at 0~T were extrapolated to the higher
voltage region for comparison with the data at 1.5~T.
Separate measurements in 0~T at the higher voltages were used
to confirm that the extrapolations were reasonable.
For signals of three or more photoelectrons delivered to the R3386 in the
absence of a magnetic field, a clear peak was
visible. 
The fractional width of the signal was found to be
equal to $(1.10\pm 0.05)N^{-{1\over 2}}$.
In the 1.5~T field the gain was found to be reduced
by a factor of $381 \pm 65$.
This corresponds to a mean loss of $\sim$31\% in multiplication
at each of the sixteen stages of the R3386.
The width of the signal was measured to be
$(1.25\pm 0.05)N^{-{1\over 2}}$.
The responses to a mean signal of 10 photoelectrons are shown in 
Figure~\ref{fig:pmt} for magnetic fields of 0~T and 1.5~T.

We conclude that the R3386 is suitable for use in our
detector but that it is important to take into account
the broadening of the signal distribution in predicting
identification efficiencies.
In our detector simulation the photomultiplier was modeled
by broadening the delivered photon signal 
by 35\%. 
This is somewhat wider than what we have observed, and should
therefore give a conservative assessment of feasibility
for this detector.

\subsection{Electronics}
The electronics to read out this detector needs to
satisfy only modest demands.
The number of channels is low, 200 or less, so that cost is
not an issue.
Resolutions of $\sim1$~ns on timing and 5\% on pulse height
are adequate, and the required dynamic range is three decades.

If the system is to operate in a high magnetic field,
preamplification of the phototube signal is required, as
the gain of the fine mesh PMT may be as low as $10^4$.
For a gain of $10^4$, a single photoelectron will result in signal charge
of 1.6~fC on the anode.
If this charge is spread over 5~ns, it gives a current amplitude 
of 0.32$\mu$A. 
A preamplifier gain of 30 mV/$\mu$A is needed to translate this
into a 10~mV signal.
The noise at the output should be less than 1~mV. 
These criteria are nearly satisfied by 
the LeCroy TRA4402, which provides a gain of 25 mV/$\mu$A with an
input noise $ <$ 65 nA RMS (1.6 mV at the output). 
There are other
amplifiers which have input noise current more than a factor of 3
lower than this \cite{amp}, which should easily meet the requirements.

\section{Prototype Tests}
\label{prototype}

\subsection{Optics}

To explore the design parameters and to verify the simulation of
this complicated optical system, we set up a model of a tubular
module with a laser configured to simulate \v{C}erenkov radiation.
The model tube was half length, 10~cm in diameter, and lined with aluminized
Mylar D-1 overcoated with MgF$_2$\cite{hyper}.
The simulated radiation, an annular ``cone'' of laser light, 
entered the tube through one of several ports along its length.
The light was produced with an opening angle of 7.9$^\circ$, 
to simulate the radiation from a 1~GeV/c pion.

\begin{figure}
\centerline{\psfig{figure=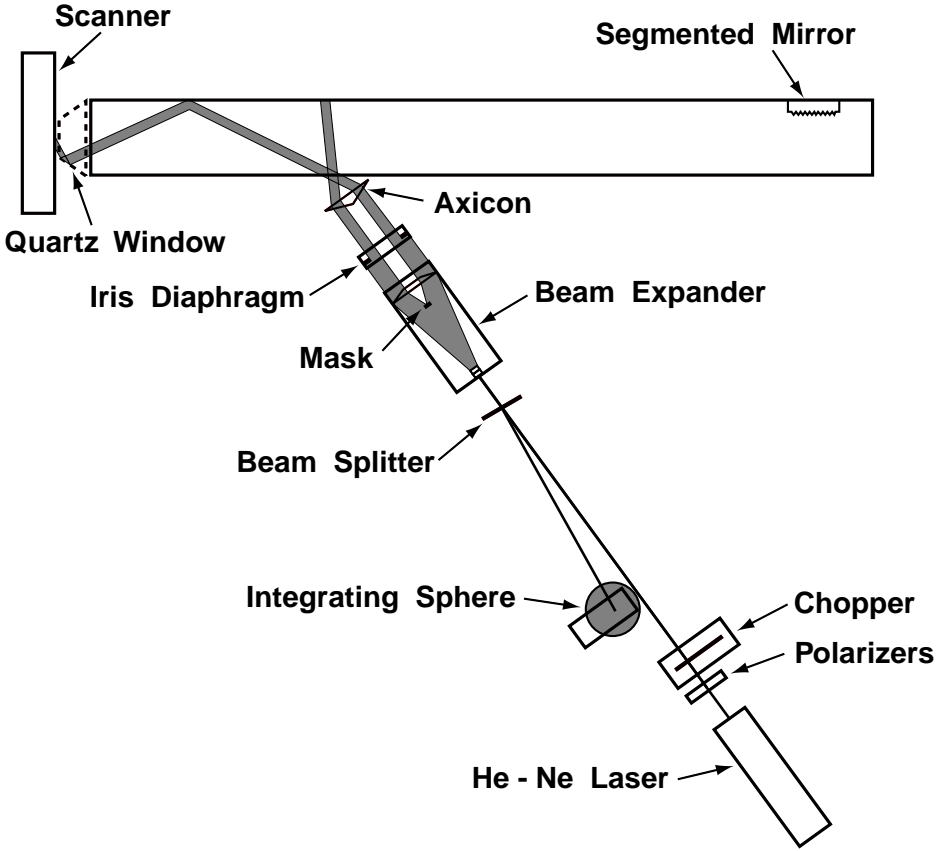,width=5.0in}}
\caption{Schematic diagram of optical test configuration.
}
\label{fig:laser}
\end{figure}

To produce the conical distribution, the monochromatic beam from a
green HeNe laser was first transported through a beam expander 
with adjustable mask and iris diaphragm to 
generate an annular beam of parallel rays.
This beam was then directed onto an axicon, which 
transformed the beam into an annular cone of light.
The location for the origin of the simulated radiation 
was adjusted by varying the mask position in the beam expander, 
and the path length over which the radiation
was emitted was set by the iris diaphragm. 
This radiation pattern differed from real \v{C}erenkov light 
in two ways: it did not have the correct spectral
distribution or polarization. 
A schematic of the beam and
apparatus is shown in Figure \ref{fig:laser}. 

One end of the tube was
equipped with an array of photodiodes attached to a movable stage that could
scan the entire area of the end of the tube. 
The source intensity was measured by shining the cone directly onto the
photodiode array.
The transport efficiency is then the ratio of light intensity detected 
at the end of the tube to the intensity of the source.
To correct for temporal fluctuations in laser power, a fraction of
the beam was extracted via a pellicle beam splitter and monitored during all measurements.
To measure its intensity in a manner insensitive to spatial fluctuations,
the extracted beam was directed into an integrating sphere, where it
was diffusely scattered from an inner wall coated with highly reflective
white paint and detected in a photodiode placed out of the beam's path.

\begin{figure}
\centerline{\psfig{figure=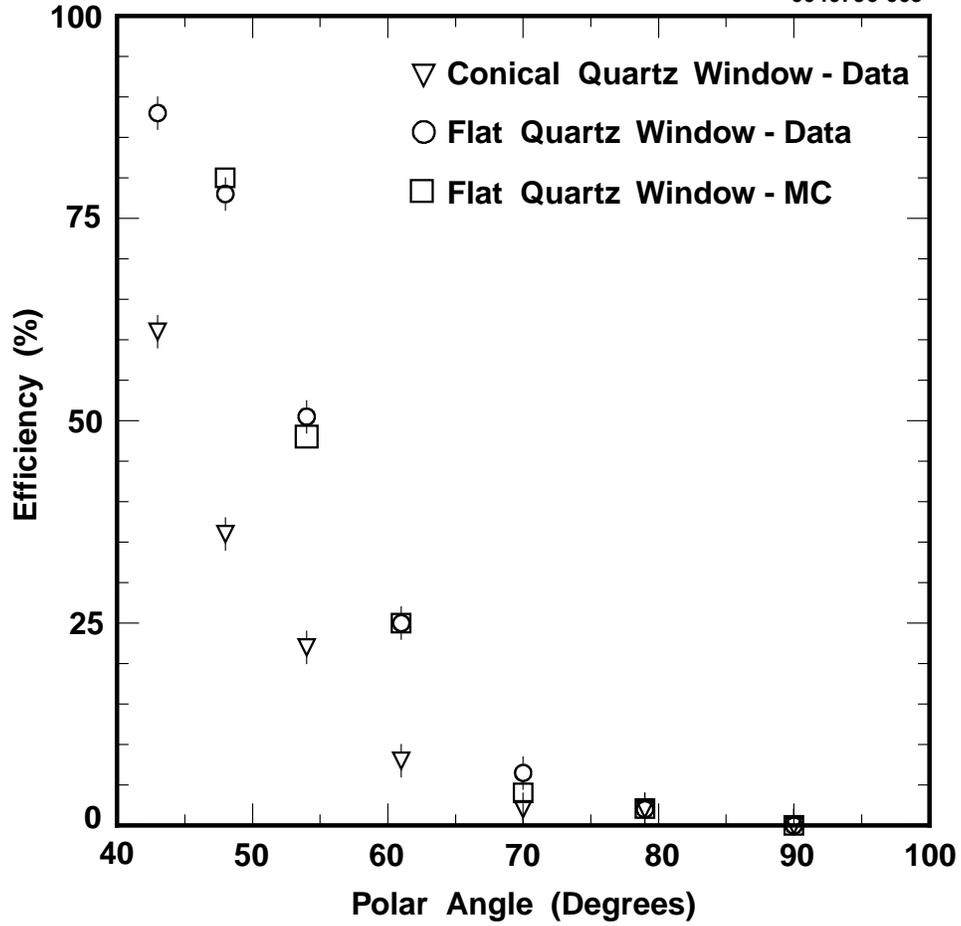,width=5.in}}
\caption{Photon transport efficiencies for green light
with plain cylindrical reflector
at large track polar angles.
Results for light
collected over the entire area of a quartz disk end window
(10~cm diameter)
show good agreement between data (squares) and Monte Carlo 
prediction (circles). 
In a real detector the photosensitive area will be no larger
than 6.4~cm diameter, so the yields for the plain disk will be 40\% of those
shown.
Also shown is the result from data taken using a
mirrored conical quartz window (triangles), where light is
collected over a  6.4~cm diameter area.
}
\label{fig:transport}
\end{figure}

In our first test, we injected the simulated \v{C}erenkov light 
at several angles, each at the position along the tube corresponding
to that expected from particles in an actual colliding beam experiment.
A 3.0~cm thick plain quartz disk was placed before the detector array to 
simulate a thick transparent endplate.
The light intensity was measured 
over the entire cross section of the tube, and the results were compared to
expectations from the simulation. 
The measured transport efficiency as a function of angle
is shown in Figure~\ref{fig:transport}, with predictions calculated by 
the same Monte Carlo simulation used to design the faceted mirror.
The agreement between the measured and simulated efficiencies is
excellent.
It varies from zero for a polar
angle of 90$^\circ$ to 88\% for incidence at 43$^\circ$, positions 
corresponding to the center and end of the tube, respectively. 

In a detector module the efficiency is further reduced by the fact that
even the larger  PMT photocathode (6.4~cm diameter) covers only 
$\sim$40\% of the 
total cross sectional area of the tube.
Thus, with a plain window and the large photocathode, only
about 40\% of the photons reaching the window would strike the 
photocathode.
This yield can be raised to around 65\% by using the conical
window discussed earlier.
Data taken with the conical window, shown as triangles in 
Figure~\ref{fig:transport}, are in agreement with this prediction.

Finally, several 5$\times$5~cm$^2$ prototype sections of the faceted 
mirror were produced commercially\cite{sollid}. 
A single section, designed for placement at 
polar angles near 90$^\circ$, was inserted near the far end of the tube.
A particle at 90$^\circ$ was simulated and the delivered light measured
with each of the two windows in place. 
The transport efficiency was found to be (22$\pm$2)\% for the plain window.
The simulation predicts a value of (23$\pm$1)\%,
in good agreement with the measurement.
These values imply an efficiency of around 8\% for delivery to a
6.4~cm diameter photocathode.
For the conical window the corresponding efficiency was 
measured to be (13$\pm$2)\%.
In a detector module, photons produced by
tracks incident at a 90$^\circ$ polar angle are reflected equally by
the faceted mirrors  to both ends of the tube, so the total transport
efficiency with the conical window will be 26\%.

\subsection{Detector module}
\begin{figure}
\centerline{\psfig{figure=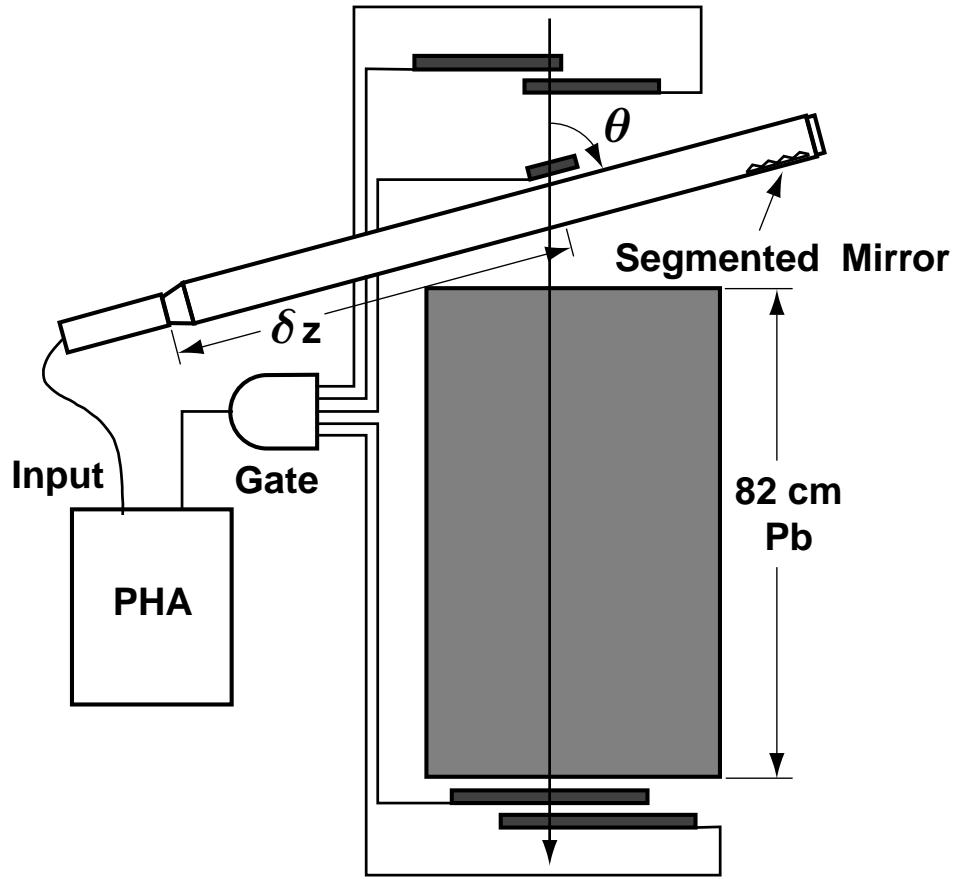,height=5.0in}}
\caption{Configuration for cosmic ray tests.  $\theta$ and
$\delta z$ are adjustable.}
\label{fig:crtower}
\end{figure}

To verify several critical aspects of the design, we constructed
one half-length module to serve as a prototype (Figure~\ref{fig:crtower}).
A 1~m long, 10~cm diameter stainless steel tube was lined with
aluminized Mylar D-1 overcoated with MgF$_2$ \cite{hyper}.
Two prototype sections of the faceted mirror were placed inside the tube
near one end, corresponding to the center of a full-length
module, and the endplate was blackened to eliminate extraneous reflections.
Placed at the other end of the tube was a Hamamatsu R4143Q photomultiplier
with quartz window,\footnote{This PMT is not what would be used in 
a final design, since it does not have the fine-mesh dynode structure.} 
viewing the pressurized gas through the mirrored conical quartz window,
described earlier.
The PMT was optically coupled to the quartz window using
General Electric Viscasil.

To study the response of this prototype module to energetic particles, 
we constructed a cosmic ray telescope, using plastic scintillators
to trigger on muons with momentum above 1~GeV/c.
Three small scintillators above the module insured that
the cosmic rays passed directly through the desired region of
the gas.
An 82~cm thick absorber of lead was placed directly below the prototype,
and two larger scintillators were placed under the lead.
To trigger all five scintillators in coincidence,
a cosmic ray muon would need not only to pass through the
prototype at the desired position and angle, but also to have
sufficient momentum to penetrate the lead, at least 1.0~GeV/c.
The prototype was 
supported on a U-channel which pivoted about a steel rod such
that its angle with respect to
the vertical axis and its position along the tube axis could be
easily modified.
A schematic view of the cosmic ray test configuration is shown in 
Figure~\ref{fig:crtower}.
The photomultiplier signal from the prototype was amplified by a 
Hewlett Packard
5582A linear amplifier and delivered to a Canberra 750Plus series
pulse height analyzer.
The gate for the signal was 4~$\mu$s wide, generated in
an EG\&G 416A gate generator by a five-fold coincidence
of the scintillators.
The measured signal was calibrated by operating the photomultiplier
with its face covered, such that the observed signal corresponded to
single photoelectron noise. 

The \v{C}erenkov light yields were studied for 
cosmic rays at many positions along the length 
of the module.
In one set of readings, the incidence angle and position were
varied simultaneously to reproduce those expected of particles
entering the actual detector.
This set consisted of the following combinations of polar angle ($\theta$) 
and distance from the window ($\delta z$): 
(a) $\theta =51.0^{\circ}$, $\delta z$=35~cm,
(b) $\theta =58.5^{\circ}$, $\delta z$=50~cm,
(c) $\theta =66.8^{\circ}$, $\delta z$=65~cm,
and (d) $\theta =73.0^{\circ}$, $\delta z$=75~cm.
In the final
detector, there would be a faceted mirror module at each of the settings 
(b), (c) and (d), but in this test none was employed.

These test configurations were duplicated in a Monte Carlo simulation, where
particles penetrate the prototype and generate \v{C}erenkov photons 
at the expected rate, distributed at the expected angles and energies.
Each photon then propagates through the gas volume and is absorbed or 
reflected at each mirrored surface encountered with a
wavelength-dependent probability based on the measured
reflectivity.
A photon reaching the front face of the end window may be either 
reflected from the quartz or transmitted.
After entering  the quartz, a photon is reflected  back out the front face, 
absorbed by the mirrored wall, or transmitted to  the photocathode. 
Once the photon enters the photocathode, the manufacturer's tabulation of
its quantum efficiency is used to determine whether a photoelectron is
produced.

\begin{figure}
\centerline{\psfig{figure=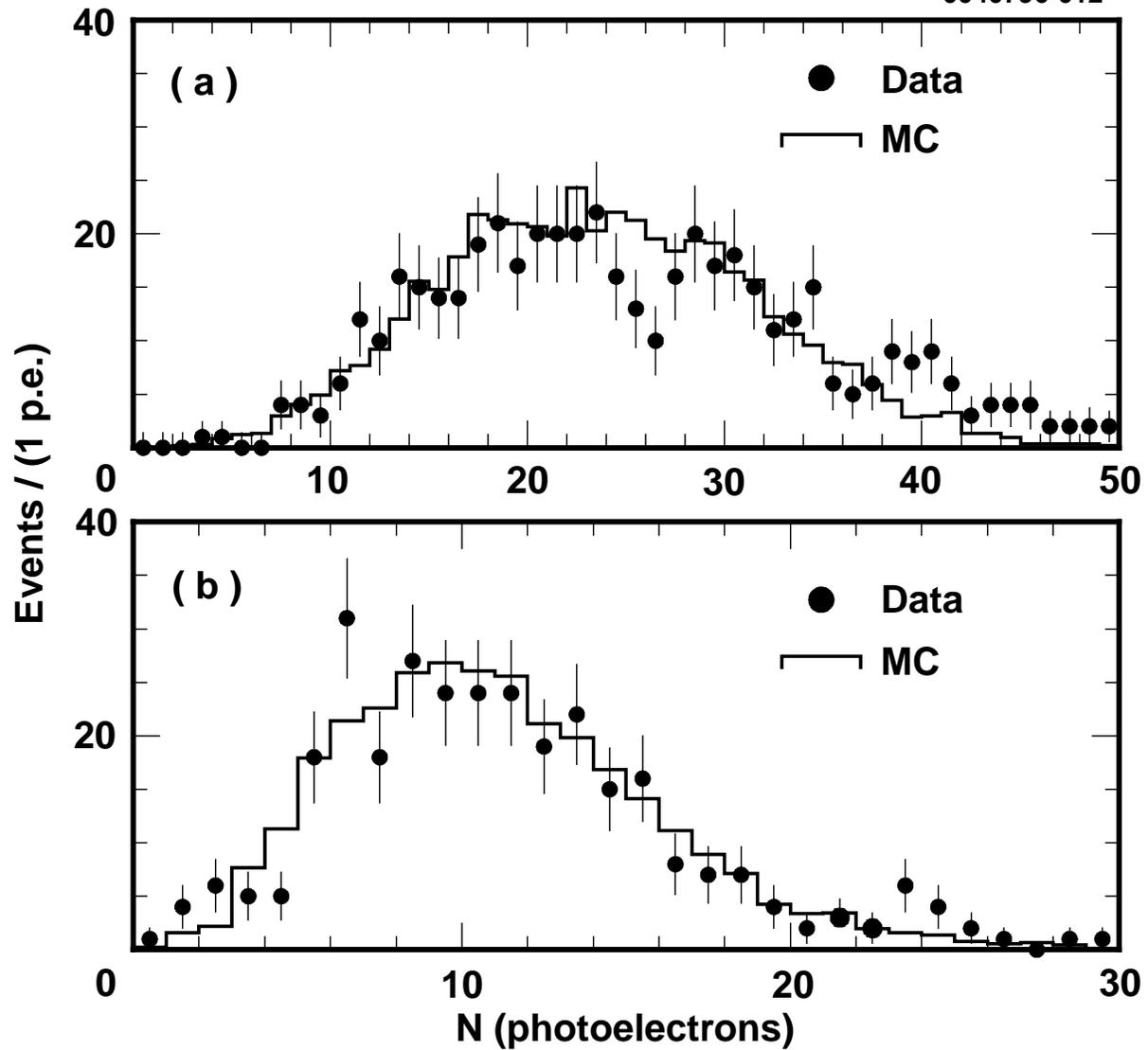,height=6.in}}
\caption{
Photoelectron yields for cosmic rays entering the prototype module at
angles which correspond to a polar angle of (a) 51$^{\circ}$, and (b)
58.5$^{\circ}$.  The points with error bars are data, and the solid histogram
is the Monte Carlo prediction.
In the design for CLEO III, the light from a
track entering at an angle of 58$^{\circ}$
would encounter a faceted mirror. 
In this test there was no faceted
mirror, thus giving a lower light yield.
}
\label{fig:data1}
\end{figure}

The data were compared with the  Monte Carlo predictions,
and simulation parameters were adjusted until agreement was achieved.
We found that excellent agreement was obtained if  the reflectivity 
of the liner in the simulation 
was adjusted upward by approximately 1\% over all wavelengths.
Since the reflectivity used originally as input to the simulation was
measured on an earlier sample than the liner in the prototype, 
it is possible that an improvement was made.
Figure~\ref{fig:data1} shows one set of data with its
prediction from the tuned simulation. 

After verifying agreement between the simulation and data without 
the faceted mirror, 
we aligned the prototype module horizontally, in an orientation 
corresponding to a polar angle of 90$^{\circ}$.  
For each of several positions, a faceted mirror section was
placed such that \v{C}erenkov photons would strike it and be directed
toward the PMT.
The mirror and trigger were set up such that
cosmic rays entered at average distances of 45~cm, 57~cm, 73~cm,
and 90~cm from the PMT.  
It is important to note that only the 90~cm position corresponds 
to a physical situation if the module were installed around 
an interaction region.  

In this configuration, because the area of a faceted mirror section is
comparable to that of the trigger counters, the raw data include events
where most of the light misses the faceted section.
To account for such events, each test was repeated with the mirror
section covered with black paper, and the resulting spectrum
was normalized to the number of triggers and subtracted directly 
from the raw data collected with the mirror section uncovered.
The results are in good agreement with our simulation, if the
reflectivity of the mirror sections is taken to be
about 0.65 times that of the reflective liner. 
The sections used here were the first replicas 
to be produced from a master. 
It is expected that with a better
understanding of the replication process, the reflectivity of the
faceted mirrors would increase substantially.

\begin{figure}
 \centerline{\psfig{figure=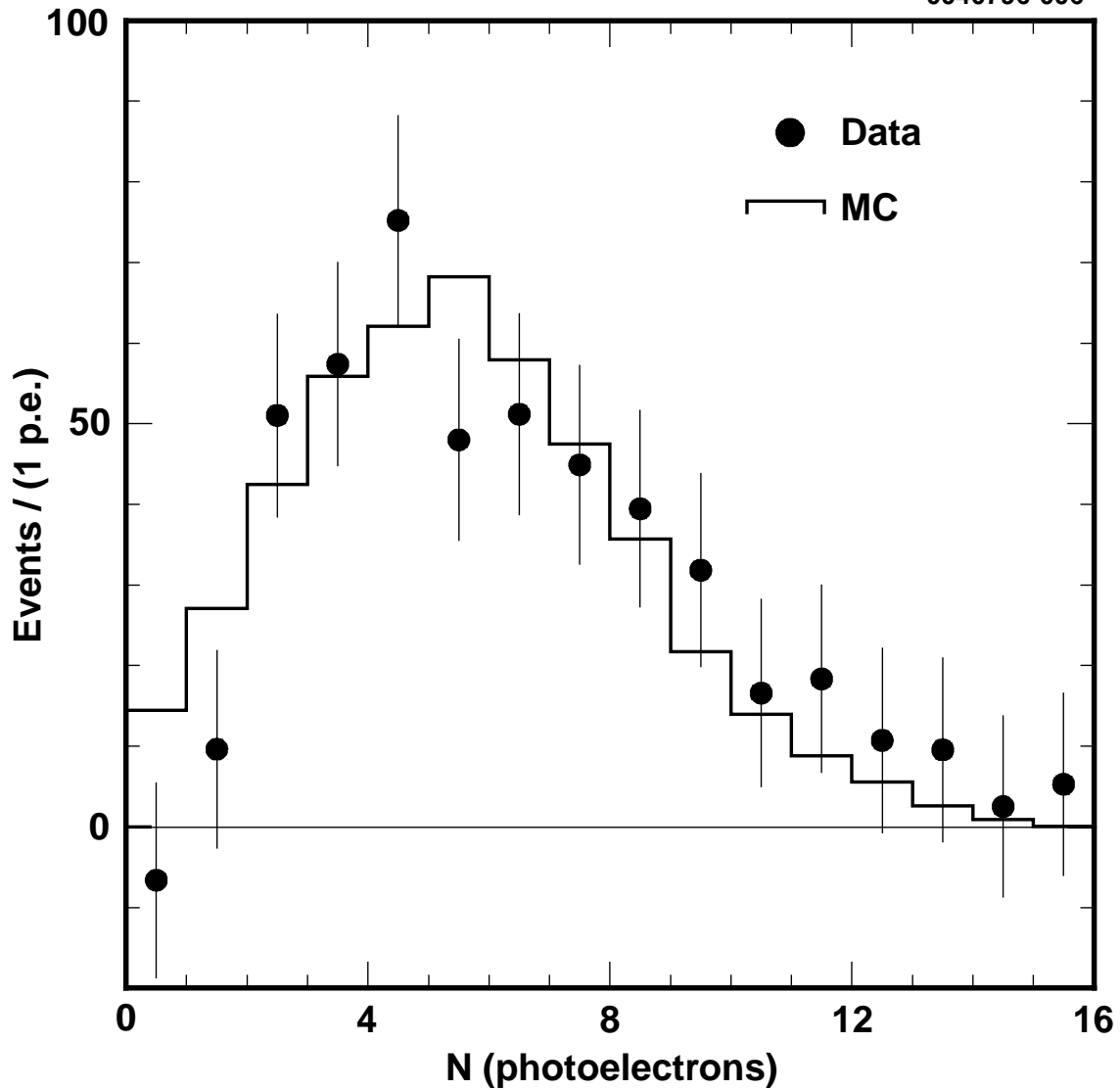,height=6.in}}
\caption{Photoelectron yield in the prototype.
Cosmic rays enter at an angle
which corresponds to a polar angle of 90$^{\circ}$, at a 
distance of 90~cm from
the photo multiplier tube, and strike a faceted mirror. 
The points with error bars are
data, and the solid histogram is the Monte Carlo prediction.
In the final system, this yield would be approximately doubled, as
the module would be approximately twice as long and
instrumented with PMT's at both ends.
}
\label{fig:data2}
\end{figure}

Results for the 90~cm position are shown in Figure~\ref{fig:data2}.
The agreement between data and Monte Carlo is excellent, giving us
confidence that the simulation is a
reasonable representation of the prototype and that the
critical parameters of this system are well understood.
In the final system, the light
collected from charged particles penetrating near the center
of the detector module would be approximately doubled, as
the module would be approximately twice as long and
would be instrumented with PMT's at both ends.

\section{Simulation Studies for CLEO~III}
\label{monte}

To evaluate the effectiveness of the full detector system 
for physics results,
the Monte Carlo simulation described in the previous section was
incorporated into the CLEO III GEANT-based \cite{geant} package.
It was then used to study two specific physics problems in 
$e^+e^-$ annihilation events in the $\Upsilon$ region:
(1) separation of modes $B \rightarrow \pi\pi$ and $B \rightarrow
K \pi$, and (2) extraction of $D^0 \rightarrow \pi l\nu$ from a
large background of $D^0 \rightarrow K l\nu$.

This simulation includes track
curvature due to the magnetic field, interactions at the various
detector interfaces, splashback from interactions in the CsI
electromagnetic calorimeter, and the performance of the fine mesh PMT in a
magnetic field. 
The signal for each module is derived by summing the outputs 
from both PMT's.
Since a track almost always traverses
two modules, the net signal is the sum of outputs from two to four
PMT's.
The noise rate of the fine mesh PMT is of the order of 1~KHz above the 
0.5~p.e. level, and the event gate in CLEO is estimated to be in the 100~ns
range; this implies that there is no significant noise contribution
when the outputs of more than one PMT are summed.

\begin{figure}
\centerline{\psfig{figure=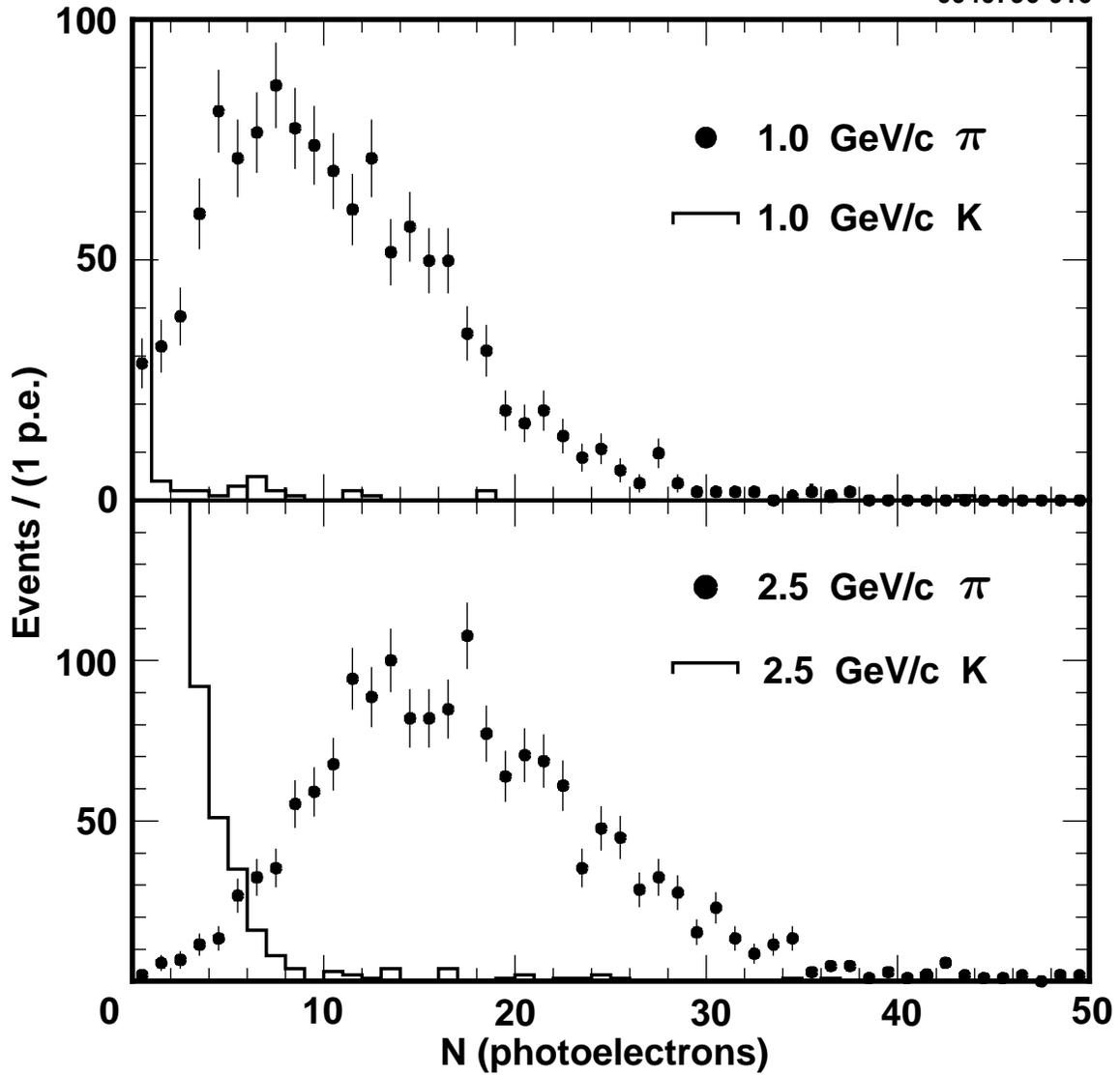,height=6.in}}
\caption{Monte Carlo prediction of signal distributions
for pions and kaons with momenta of
1.0~GeV/c and 2.5~GeV/c incident on the detector with polar 
angle between 85$^{\circ}$ and 95$^{\circ}$.}
\label{fig:pik25}
\end{figure}

First, single $\pi$ and $K$ tracks were generated with momenta in 
the range 0.8-3.0~GeV/c and polar angles between
45$^{\circ}$ and 135$^{\circ}$ (this angular region is set by the 
active volume of the detector). 
As an example, we show in Figure~\ref{fig:pik25} the expected
photoelectron yield for pions and kaons of momenta 1.0 and 2.5~GeV/c and polar
angle between 85$^{\circ}$ and 95$^{\circ}$. 
At 2.5~GeV/c the kaons are slightly above threshold.
Beyond the expected signal,
it is seen that a small fraction of kaons produce a substantial
false signal due to an interaction in the material before the detector,
which may produce a pion or delta ray which is above threshold.
To be identified as pions (kaons), we require the 
photoelectron yield to be above (below) a certain value.
The cut values are different for pions and kaons and are 
functions of the track momentum and polar angle.  
To reduce the rate of false identifications due to interactions,
we require alignment between
the projected track and the shower centroid as measured in the calorimeter.
The degree of alignment required is in the range 1.7$^\circ$-2.3$^\circ$ and
depends primarily on the track momentum.
It is chosen such that 80\% of pions pass it.
In Table~\ref{tab:str} we show the net pion efficiency, $\epsilon_\pi$, 
and kaon fake rate, $f_K$,
where the cuts have been optimized by maximizing $\epsilon_\pi/f_K$, which is
proportional to the ratio of signal to background (S/B).
At momenta
2.5-2.8~GeV/c, S/B ranges from about 30 to 60, depending on the
angle of the track; this translates roughly to 2.5$\sigma$ $\pi /K$
separation. 
Combining this information with measurements of specific ionization in 
the drift chamber
will increase the $\pi /K$ separation to above 3$\sigma$ \cite{dedx}.

Using the criteria developed from simulations of single $\pi$ and $K$ tracks, we
studied some specific decay modes in fully simulated events. 
The efficiencies presented below are
due only to particle identification, since all other efficiencies, such
as geometric cuts, kinematic cuts, lepton identification, and
isolation of tracks to eliminate multiple hits,  cancel to
first order for this evaluation.

We have studied the two decays $B\rightarrow \pi \pi$ and
$B\rightarrow K\pi$, where each
mode poses the most serious background for the other and the two
branching fractions are thought to be comparable.
If we choose pulse height cuts by maximizing S/B, we find that the PID
efficiency for detecting the $K \pi$ final state is $(23.2 \pm
1.0)\%$, whereas the background from the $\pi\pi$ state is $(0.3 \pm
0.1)\%$. 
If instead, we wish to detect $B\rightarrow \pi \pi$, then
the efficiency is $(22.7 \pm 1.0)\%$ and the background from the $K
\pi$ final state is $(1.2\pm 0.1)\%$. 
These efficiencies imply a rejection ratio of at least 1/20, which 
corresponds to a separation of 2.5$\sigma$.  
The S/B for the two modes are different because kaons
have a long tail at high photoelectron yields, while pions do not
have a similar tail at low yields. 
Further separation between the two
samples can be achieved by using dE/dx information \cite{dedx} from the drift
chamber and the kinematic quantity $\Delta E \equiv E_{beam} - E_B$. 

We also studied the decay $D^0\rightarrow \pi^-l^+\nu$, a Cabbibo-suppressed
decay where there is a large background from the unsuppressed
decay  $D^0\rightarrow K^-l^+\nu$.
In this case, cuts were optimized by maximizing S$^2$/(S+B),
where the $\pi$ mode was assumed to have a rate 15 times
smaller than the $K$ mode.
We selected $D$ mesons produced with momenta above 1.9~GeV/c, 
and with $\pi$ or $K$ polar angle between 45$^{\circ}$ and 135$^{\circ}$. 
The resulting efficiencies for
detecting the pion in $D^0\rightarrow\pi^-l^+\nu$ and fake rates from kaons in 
$D^0\rightarrow K^-l^+\nu$ are presented in Table~\ref{tab:semil}. 
These results indicate a S/B of about 20. 
It is clear that, even after allowing for the higher rate for $D^0\rightarrow
K^-l^+\nu$, the final sample is enriched with the signal decay. 
This is
much better than the current CLEO II performance\cite{dsemil}, where
the final sample is dominated by background from $K^-l^+\nu$ events. 
Additional signal enhancement may be achieved through kinematic requirements
and measurements of specific ionization\cite{dedx}.
     
\section{Conclusion}

It has been demonstrated that a simple, compact gas threshold \v{C}erenkov 
counter which utilizes commercially produced photodetectors
can provide hadron identification in the momentum region 1-3~GeV/c.
The design described here can be implemented as is, with the
development of a mechanical design.
Possible improvements requiring some further development
include better quality reflective liners and faceted mirrors, 
optimized design of the quartz window,  and use of low-$Z$ materials
such as carbon fiber or beryllium in the tubes to reduce the 
thickness of material. 
In addition,
continuing improvements in photodetection technology are beginning
to yield devices with better operating characteristics, {\it e.g.},
higher quantum efficiency, than the fine mesh PMT studied here\cite{private}.  

Although this detector was developed for implementation in the
CLEO~III detector, it could be used in a much broader set of applications.
Because of its speed and the low number of readout channels, it can
be used in most settings with moderately high rate and low 
particle multiplicities.
By using multiple layers and varying gas pressures,
this design can be applied to all particles with momenta above 1~GeV/c, up 
to 10~GeV and above.
These are difficult regions in which to achieve hadron identification,
and the detector presented here does so simply and at moderate 
cost.

\section{Acknowledgments}

Thanks are due for assistance provided by colleagues
in the CLEO collaboration, to Brian Heltsley for assistance with 
incorporating our Monte Carlo into the CLEO III GEANT environment,
to Jim Alexander for providing funds for materials from his NSF-PYI award, 
and to the High Energy Physics group at the University of California,
Santa Barbara, for use of equipment.
We thank Karl Berkelman and the Laboratory
of Nuclear Studies, Cornell University, for providing laboratory space
and material for conducting this research. 
Jim McCormick of Hamamatsu
Corp.~is gratefully acknowledged for providing us with a fine mesh
photomultiplier tube for test purposes.
Jon E. Sollid of Sollid Optics, Inc., is acknowledged for
providing valuable advice on the design and for finding sources
for many of the optical components critical to our tests.

   This research was funded by the U.S. National Science Foundation, 
the U.S. Department of 
Energy, and Vanderbilt University.
%
%

%

%
%


\begin{table}
\caption{The orientations of the mirror facets. $\alpha$ and $\beta$
are defined in Figure~\ref{fig:grating}}
\begin{tabular}{cccc}
$|z|$ interval (m) & polar angle (degrees) & $\alpha $ (degrees) 
& $\beta $ (degrees) \\ \hline
{0.00 - 0.05} & 88.4 & {37.5} & {37.5} \\ 
{0.05 - 0.11} & 84.9 & {32.0} & {41.0} \\ 
{0.11 - 0.15} & 81.8 & {29.0} & {42.0} \\ 
{0.15 - 0.21} & 78.7 & {26.0} & {43.0} \\ 
{0.21 - 0.26} & 75.4 & {23.0} & {44.0} \\ 
{0.26 - 0.32} & 72.1 & {20.0} & {45.0} \\ 
{0.32 - 0.38} & 68.7 & {18.0} & {46.0} \\ 
{0.38 - 0.45} & 65.2 & {16.0} & {47.0} \\ 
{0.45 - 0.53} & 61.4 & {14.0} & {48.0} \\ 
{0.53 - 0.61} & 57.7 & {12.0} & {49.0} \\ 
\end{tabular}
 \label{tab:angles}
\end{table}

\begin{table}
\setlength{\tabcolsep}{2.0pc}
\caption{Pion efficiencies $\epsilon_\pi$ and kaon fake rates $f_K$,
in percent, for single tracks at three different polar angles. 
Cuts have been optimized by maximizing S/B.}
\label{tab:str}
\begin{tabular}{cccc}
  Momentum (GeV/c) &  $50\pm 5^{\circ}$ & $70\pm 5^{\circ}$ 
& $90\pm 5^{\circ}$ \\
        & $\epsilon_\pi$\ \ ($f_K$) & $\epsilon_\pi$\ \ ($f_K$)
& $\epsilon_\pi$\ \ ($f_K$) \\
\hline
  0.8   & 54.0\ \ (0.65)   & 70.8\ \  (0.90)     &\ \  62.8 (1.35)  \\
  1.0   & 54.0\ \  (0.25)   & 72.1\ \  (1.40)     &\ \  61.9 (1.10)    \\
  1.2   & 59.6\ \  (0.20)   & 71.2\ \  (0.80)     &\ \  62.4 (0.80)   \\
  1.6   & 59.9\ \  (0.87)   & 70.6\ \  (1.15)     &\ \  62.7 (1.20)   \\
  2.0   & 54.6\ \  (0.65)   & 70.3\ \  (1.50)     &\ \  61.1 (0.85) \\
  2.5   & 37.6\ \  (1.05)   & 71.6\ \  (1.30)      &\ \  61.2 (0.95)  \\
  2.8   & 36.3\ \  (1.10)   & 42.2\ \  (1.55)      &\ \  28.3 (0.90) \\
  3.0   & 36.5\ \  (2.15)   & 32.1\ \  (1.30)      &\ \  33.5 (3.50) \\
\end{tabular}
\end{table}

\begin{table}
\setlength{\tabcolsep}{2.0pc}
\caption{Efficiencies for $D^0 \rightarrow \pi^- l^+\nu$, fake rates from
$D^0 \rightarrow K^- l^+\nu$}
\label{tab:semil}
\begin{tabular}{ccc}
  Track momentum (GeV/c) & Efficiency, $\pi l\nu$ & Fake rate, $Kl\nu$ \\ 
\hline
  0.8-1.0  & 70.7\%  & 4.6\%  \\
  1.0-1.2      & 67.3\%  & 4.5\%   \\
  1.2-1.6      & 71.3\%  & 2.6\%       \\
  1.6-2.0      & 74.3\%  & 4.1\%       \\
  2.0-2.8      & 71.1\%  & 2.9\%     \\
  2.8-3.0      & 46.8\%  & 4.4\%     \\
\hline
 0.8-3.0   & $(70.5\pm 1.3)\%$ & $(3.6 \pm 0.3)\%$ \\
\end{tabular}
\end{table}



\begin{references}
\bibitem{cleoiii} The CLEO III Detector Status Report, CBX95-96,
Laboratory of Nuclear Studies, Cornell University.

\bibitem{aerogel}
T. Hasegawa {\it et al.}, Nucl. Inst. and Meth. {\bf A342} (1994) 383;
I. Adachi {\it et al.}, Nucl. Inst. and Meth. {\bf A355} (1995) 386.

\bibitem{Atkinson} J.~H.~Atkinson and V.~Perez-Mendez, Rev. Sci.
Inst. {\bf 30} (1959) 864.

\bibitem{Garwin} E.~L.~Garwin and A.~Roder, Nucl. Inst. and Meth. 
{\bf 93} (1971) 593.

\bibitem{foot} Since the operating pressure is very close to the
liquification point of SF$_6$, transmission immediately after filling was
poor.  Once stable,
transmission was observed to remain constant over periods of greater
than one week.

\bibitem{Tomkiewicz} Y.~Tomkiewicz and E.~L.~Garwin, Nucl. Inst. and Meth. 
{\bf 114} (1974) 413.

\bibitem{hyper} The liner used in the tests was coated by Hyperfine Inc., 
Colorado.

\bibitem{width}
According to our simulation, the facet strips may be as wide as 3~mm 
without undue loss of signal.  
The larger width is desirable for reducing the cost associated 
with fabricating master forms for the mirrors but
increases the net thickness of the mirror.
The 1~mm width in our design was chosen as a compromise between thickness
and cost.

\bibitem{fmpmt}
A. Sawaki {\it et al.}, IEEE Trans. Nucl. Sci. {\bf NS-31, 1} (1984) 442;
M. D. Rousseau {\it et al.}, IEEE Trans. Nucl. Sci. {\bf NS-30, 1} (1983) 479;
H. Kume {\it et al.}, IEEE Trans. Nucl. Sci. {\bf NS-32, 1} (1985) 355;
S. Suzuki {\it et al.}, IEEE Trans. Nucl. Sci. {\bf NS-33, 1} (1986) 377.

\bibitem{enomoto}
R. Enomoto {\it et al.}, Nucl. Inst. and Meth. {\bf A332} (1993) 129.

\bibitem{amp} P. Jarron and M. Goyot, Nucl. Inst. and Meth. 
{\bf 226} (1984) 156.

\bibitem{sollid} The faceted mirrors were constructed and coated by 
Sollid Optics, Inc., New Mexico.

\bibitem {geant}
R. Brun {\it et al.}, GEANT 3.15, CERN DD/EE/84-1.

\bibitem{dedx} In CLEO II pions and kaons of momenta less than 0.8 GeV/c can
be separated by better than 3.0$\sigma$, and those above 
2 GeV/c or so can be separated by 1.5-2.0$\sigma$. 
It is expected that CLEO III performance will be similar.

\bibitem{dsemil} 
F. Butler {\it et al.} (CLEO), Phys. Rev. {\bf D 52} (1995) 2656.

\bibitem{private} Private communication with J. McCormick, Hamamatsu Corp.

\end{references}
\end{document}